\documentclass[twoside,leqno,twocolumn]{article}  
\usepackage{ltexpprt} 
\usepackage{graphicx}
\begin{document}

\title{\Large {Artificial Inflation: The True Story of Trends in Sina Weibo}}


\author{Louis Lei Yu\small{*}
\and 
Sitaram Asur\small{*}
\and
Bernardo A. Huberman\thanks{Social Computing Lab, HP Labs. Email:{\tt \{louis.yu, sitaram.asur, bernardo.huberman\}@hp.com}}}
       
\date{}

\maketitle


\begin{abstract} 
There has been a tremendous rise in the growth of online social networks all over the world in recent years. This has facilitated users to generate a large amount of real-time content at an incessant rate, all competing with each other to attract enough attention and become trends. While Western online social networks such as Twitter have been well studied, characteristics of the popular Chinese microblogging network Sina Weibo have not been. In this paper, we analyze in detail the temporal aspect of trends and trend-setters in Sina Weibo, constrasting it with earlier observations on Twitter. First, we look at the formation, persistence and decay of trends and examine the key topics that trend in Sina Weibo. One of our key findings is that retweets are much more common in Sina Weibo and contribute a lot to creating trends. When we look closer, we observe that a large percentage of trends in Sina Weibo are due to the continuous retweets of a small amount of fraudulent accounts. These fake accounts are set up to artificially inflate certain posts causing them to shoot up into Sina Weibo's trending list, which are in turn displayed as the most popular topics to users.
\end{abstract}
\section{Introduction}
In the past few years, social media services as well as the users who subscribe to them, have grown at a phenomenal rate. 
This immense growth has been witnessed all over the world with millions of people of different backgrounds using these services on a daily basis to 
communicate, create and share content on an enormous scale. This widespread generation and consumption of content has created an extremely complex and competitive online environment where different types of content compete with each other for the attention of users. Thus it is very interesting to study how certain types of content such as a viral video, a news article, or an illustrative picture,  manage to attract more attention than others, thus bubbling to the top in terms of popularity.  Through their visibility, these  popular items and topics contribute to the collective awareness reflecting what is considered important. This can also be powerful enough to affect the public agenda of the community.

There have been studies on trends and trend-setters in Western online social media \cite{Asur2011} \cite{Huberman}. In this paper, we examine in detail a significantly less-studied but equally fascinating online environment: Chinese social media, in particular, Sina Weibo: China's biggest microblogging network.

Over the years there have been news reports on various Internet phenomena in China, from the surfacing of certain viral videos  to the spreading of rumors \cite{Jin} to the so called ``human flesh search engines'' \cite{flesh}  \footnote{is a primarily Chinese Internet phenomenon of massive search using online media such as blogs and forums \cite{flesh}.} in Chinese online social networks. These stories seem to suggest that many events happening in Chinese online social networks are unique products of China's culture and social environment.

Due to the vast global connectivity provided by social media, netizens \footnote{A netizen is a person actively involved in online communities \cite{Netizen}.}  all over the world are now connected to each other like never before. This means that they can now share and exchange ideas with ease. It could be argued that this means the manner in which the sharing occurs is similar across countries.
However, China's unique cultural and social environment  suggests that the way individuals share ideas might be different than that in Western societies. For example, the age of Internet users in China is a lot younger, they may respond to different types of content than Internet users in Western societies. The number of Internet users in China is larger than that in the U.S, and the majority of users lives in large urban cities. One would expect that the way these users share information can be even more chaotic. An important question to ask is to what extent would topics have to compete with each other in order to capture users' attention in this dynamic environment. Furthermore, it is known that the information shared between individuals in Chinese social media is monitored \cite{Tai}. Hence another interesting question to ask is what types of content would netizens respond to and what kind of popular topics would emerge under such circumstances. 

Given the above questions, we present an analysis on the evolution of trends in Sina Weibo. We have monitored the evolution of the top trending keywords in Sina Weibo for 30 days. First, we analyze the model of growth of these trends and examine the persistance of these topics over time.
We investigate if topics initially ranked higher tend to stay in the top 50 trending list longer. 
Subsequently, by analyzing the timestamps of posts containing the keywords, we look at the propagation and decaying process of the trends in Sina Weibo and compare it to earlier observations on Twitter~\cite{Asur2011}. We establish that retweets play a greater role in Sina Weibo than on Twitter, contributing more to the generation and persistance of trends. When we examine the retweets in detail, we make an important discovery. 
We observe that many of the trending keywords in Sina Weibo are heavily manipulated and controlled by certain fraudulent accounts that are setup for this purpose. We found significant evidence suggesting that a large percentage of trends in Sina Weibo are actually due to artificial inflation by these fake users, thus making certain posts trend and be more visible to other users.
The users we identified as fraudulent were 1.08\% of the total users sampled, but they were responsible for 49\% of the total retweets (32\% of the total tweets). We evaluate some methods to identify these fake users and demonstrate that by removing the tweets associated with the fake users, the evolution of the tweets containing the trending keywords follow the same persistent and decaying process as the one in Twitter. 

The rest of the paper is organized as follows: in Section 2 we give some related work. In Section 3 we give some background information on the development of the Internet and online social networks in China. We analyze the trends and trend-setters on Sina Weibo in Section 4 and finally in Section 5, we conclude and discuss our findings.

\section{Related Work}


\subsection{The Study of Chinese Online Social Networks}
Jin \cite{Jin} has studied the Chinese online Bulletin Board Systems (BBS), and provided observations on the structure and interface of Chinese BBS and the  behavioral patterns of its users.  Xin \cite{Xin} has conducted a survey on BBS's influence on the  University students in China and their behavior on Chinese BBS.  Yu and King.  \cite{yu} has looked at the adaptation of interests such as books, movies, music, events and discussion groups on Douban, an online social network and media recommendation system frequently used by the youth in China. 


\subsection{Trends and Trend-setters in Twitter }
There are various studies on trends on Twitter \cite{Huberman} \cite{Kwak} \cite{Mathioudakis} \cite{Wu2}.  Recently, Asur and others~\cite{Asur2011} have examined the growth and persistence of trending topics on Twitter. They discovered that traditional media sources are important in causing trends on twitter. Many of the top retweeted articles that formed trends on Twitter were found to arise from news sources such as the New York Times. In this work, we evaluate how the trending topics in China relate to the news media.

\subsection{Trend-setters in Sina Weibo}
Yu et al.  gave the first known study of trending topics in a Chinese online microblogging social network (Sina Weibo) \cite{Weibo}. They discovered that there are vast differences between the content that is shared on Sina Weibo and that on Twitter. In China, people tend to use microblogs to share jokes, pictures and videos and a significantly large percentage of posts are retweets. The trends that are formed are almost entirely due to the retweeting of such media content. This is contrary to what was observed on Twitter, where the trending topics have more to do with current events and the effect of retweets is not as high. 

\section{Background}
\subsection{The Internet in China}

The development of the Internet industry in China over the past decade has been impressive. According to a survey from  the China Internet Network Information Center (CNNIC), by July 2008, the number of Internet users in China has reached 253 million, surpassing the U.S. as the world's  largest Internet market \footnote{ The 21st statistics report on the Internet development in china is available (in chinese) at \textit{http://www.cnnic.cn/uploadfiles/pdf/2008/2/29/104126.pdf}}. Furthermore, the number of Internet users in China as of 2010 was reported to be 420 million.

Despite this, the fractional Internet penetration rate in China is still low.  The 2010 survey by CNNIC on the Internet development in China \footnote{The 26th statistics report on the internet development in china is available (in chinese) at \textit{http://www.cnnic.cn/uploadfiles/pdf/2010/8/24/93145.pdf}} reports that the Internet penetration rate in the rural areas of  China is on average $5.1\%$. In contrast,  the  Internet penetration rate in the urban cities of China is on average $21.6\%$. In metropolitan cities such as Beijing and Shanghai, the Internet penetration rate has reached over $45\%$, with Beijing being $46.4\%$ and  Shanghai  being $45.8\%$. According to the survey,  China's cyberspace is dominated by young urban students between the age of 18--30. 


The Government plays an important role in fostering the advance of the Internet industry in China.   According to  \textit{The Internet in China}  released by  the Information Office of the State Council of China\footnote{``The Internet in China'' by the Information Office of the State Council of the People's Republic of China is available at \textit{http://www.scio.gov.cn/zxbd/wz/201006/t667385.htm}}: 

 \begin{quote}  The Chinese government attaches great importance to protecting the safe flow of Internet information, actively guides people to manage websites in accordance with the law and use the Internet in a wholesome and correct way. \end{quote}

Online social networks are a major part of the Chinese Internet culture \cite{Jin}.  Netizens in China organize themselves using  forums, discussion groups, blogs, and other social networking platforms to engage in  activities such as exchanging viewpoints and sharing information \cite{Jin}.  According to \textit{The Internet in China}:
 
 \begin{quote} 
 Vigorous online ideas exchange is a major characteristic of China's Internet development, and the huge quantity of BBS posts and blog articles is far beyond that of any other country. China's websites attach great importance to providing netizens with opinion expression services, with over 80\% of them providing electronic bulletin service. In China, there are over a million BBSs and some 220 million bloggers. According to a sample survey, each day people post over three million messages via BBS, news commentary sites, blogs, etc., and over 66\% of Chinese netizens frequently place postings to discuss various topics, and to fully express their opinions and represent their interests. The new applications and services on the Internet have provided a broader scope for people to express their opinions. The newly emerging online services, including blog, microblog, video sharing and social networking websites are developing rapidly in China and provide greater convenience for Chinese citizens to communicate online. Actively participating in online information communication and content creation, netizens have greatly enriched Internet information and content.
  \end{quote}

\subsection{Sina Weibo}

Sina Weibo was  launched by the Sina corporation, China's biggest web portal, in August 2009.   According to ``Microblog Revolutionizing China's Social Business Development''  released by the Sina corporation  in October, 2011\footnote{``Microblog Revolutionizing China's Social Business Development'' by the Sina corporation and CIC is available at \textit{http://www.ciccorporate.com}},  Sina Weibo now has 250 million registered accounts and generates 90 million posts per day.

While both Twitter and Sina Weibo enable users to post messages of up to 140 characters, there are some differences in terms of the functionalities offered. We give a brief introduction on Sina Weibo's interface and functionalities. 

\subsubsection {User Profiles}

Similar to Twitter, a user profile on Sina Weibo displays the user's name, a brief description of the user, the number of followers and followees the user has, and the number of tweets the user made. A user profile also displays the user's recent tweets and retweets. 

There are three types of user accounts on Sina Weibo, regular user accounts, verified user accounts, and the expert (star) user account.  A verified user account typically represents a well known public figure or organization in China.  Sina has reported in the annual report that it has more than 60,000 verified accounts consisting of celebrities, sports stars, well known organizations (both Government and commercial) and other VIPs.  

Regular users can register to become an expert (star) user,  part of  Sina Weibo's new user  account verification program: ``DaRen'' (expert in Chinese). This program offers incentives for users to verify their personal information such as passport numbers (or ID numbers for Chinese citizens) and home addresses. Some criteria for a regular users to become an expert (star) user includes\footnote{See instructions (in Chinese) at \textit{http://club.weibo.com/}  on how to become an expert (star) user}: 1)  the user profile picture must be a photo of user himself/herself; 2) the user must provide a registered mobile number; 3)  the user account must follow at least 100 people, among them at least 30\% must be the followers of user; 4) the user accounts must have at least 100 followers.

\subsubsection{The Content of Tweets on Sina Weibo}

There is an important difference in the content of tweets between Sina Weibo and Twitter.
While Twitter users can post tweets consisting of text and links,  Sina Weibo users can post messages containing text, pictures, videos and links. 

\begin{figure} [ht]
\centering
\includegraphics[width=80mm, height=50mm] {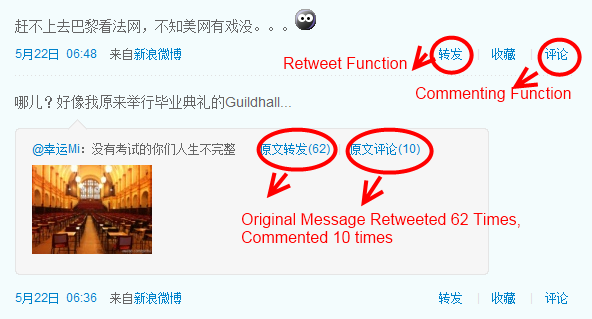}
\caption{ An Example of a Retweet on Sina Weibo (Translations of the Tweets Omitted) } \label{retweet}
\end{figure}

Twitter users can address tweets to other users and can mention others in their tweets [13].  A  common practice on Twitter is ``retweeting'',  or rebroadcasting someone else's messages to one's followers.  The equivalent of a retweet on Sina Weibo is instead shown as two amalgamated entries: the original entry and the current user's actual entry which is a commentary on the original entry. Figure \ref{retweet} illustrates an example of a retweet on Sina Weibo.

Sina Weibo  has another functionality absent from Twitter: the comment. When a Weibo user makes a comment, it is not rebroadcasted to the user's followers. Instead, it can only be accessed under the original message.

\section{Experiments and Results}
\subsection{The Trending Keywords}

Sina Weibo offers a list of 50 keywords that appear most frequently in users' tweets. They are ranked according to the frequency of appearances in the last hour. This is similar to Twitter, which also presents a constantly updated list of trending topics: keywords that are most frequently used in tweets over a period of time. We extracted these keywords over a period of 30 days and retrieved all the corresponding tweets containing these keywords from Sina Weibo.



\begin{figure} [ht]
\centering
\includegraphics[width=80mm, height=40mm] {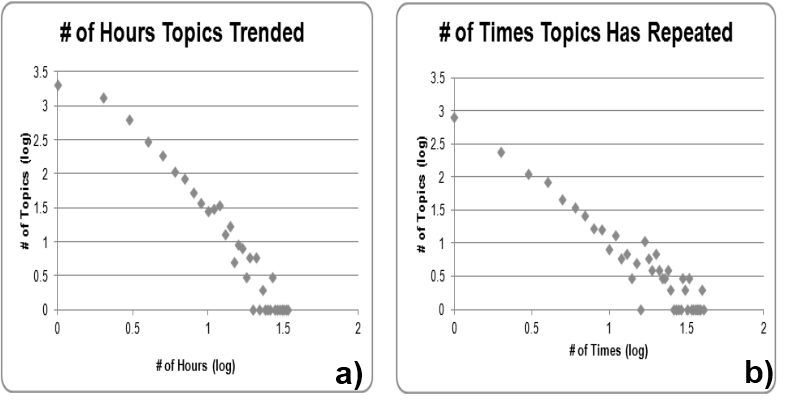}
\caption{ The distribution for the number of hours topics trended and the number of times topics reappeared } \label{power}
\end{figure}

We first monitored the hourly evolution of the top 50 keywords in the trending list for 30 days. We observed that the average time spent by each keyword in the hourly trending list is 6 hours. And, the distribution for the number of hours each topic remains on the top 50 trending list follows the power law (as shown  in Figure \ref{power} a). The distribution suggests that only a few topics exhibit long-term popularity. 

Another interesting observation is that a lot of the key words tend to disappear from the top 50 trending list after a certain hour and then later reappear. We examined the distribution for the number of time keywords have reappeared in the top 50 trending list (Figure \ref{power} b). We observe that this distribution follows the power law as well.

Both the above observations are very similar to our earlier study of trending topics on twitter~\cite{Asur2011} although the average trending time is significantly higher on Sina Weibo (on Twitter it was 20-40 minutes).

\begin{figure} [ht]
\centering
\includegraphics[width=80mm, height=50mm] {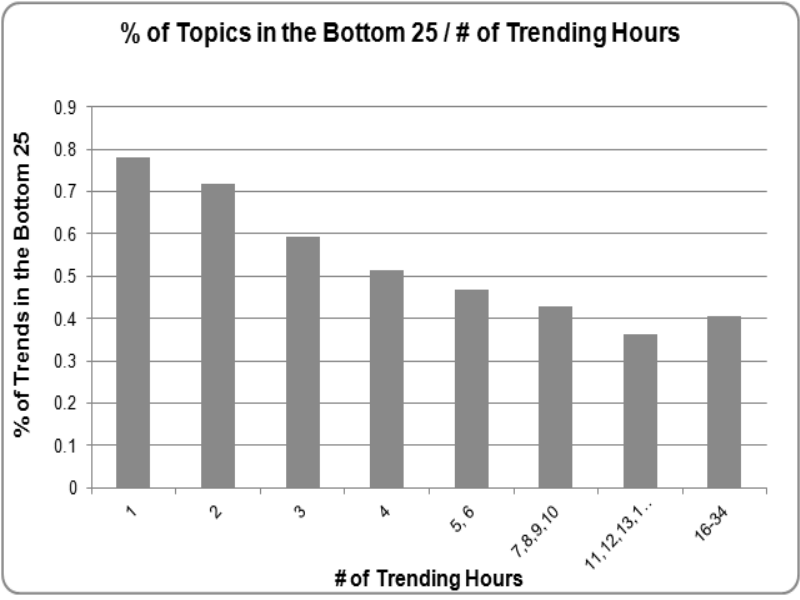}
\caption{ Percentage of the Topics Placed in the Bottom 25 Versus the Hour They Stayed in the Top 50 List} \label{stay}
\end{figure}

Following our observation that some keywords stay in the top 50 trending list longer than others, we wanted to investigate if topics that ranked higher initially tend to stay in the top 50 trending list longer.  We separated the top 50 trending keywords into two ranked sets of 25 each: the top 25 and the bottom 25. Figure \ref{stay} illustrates the plot for the percentage of topics that placed in the bottom 25 relating to the number of hours these topics stayed in the top 50 trending list. We can observe that topics that do not last are usually the ones that are in the bottom 25. On the other hand, the long-trending topics spend most of their time in the top 25, which suggests that items that become very popular are likelier to stay longer in the top 50. This intuitively means that items that attract phenomenal attention initially are not likely to dissipate quickly from people's interests.

\subsection{The Evolution of Tweets}
 
Next, we want to investigate the process of persistence and decay for the trending topics on Sina Weibo. In particular, we want to measure the distribution for the time intervals between tweets containing the trending keywords. We continuously monitored the keywords in the top 50 trending list. For each trending topic we retrieved all the tweets containing the keyword from the time the topic first appeared in the top 50 trending list until the time it disappeared. Accordingly, we monitored 811 topics over the course of 30 days. 
In total we collected 574,382 tweets from 463,231 users.  Among the 574,382 Tweets,  35\% of the tweets (202,267 tweets) are original tweets, and 65\% of the tweets (372,115 tweets) are retweets. 40.3\% of the total users (187130 users)  retweeted at least once in our sample. 


We measured the number of tweets that each topic gets in 10 minutes intervals, from the time the topic starts trending until the time it stops. From this we can sum up the 
tweet counts over time to obtain the cumulative number of tweets $N_q(t_i)$ of topic $q$ for any time frame $t_i$,
This is given as :
\begin{equation}
N_q(t_i) = \sum_{\tau = 1}^{i} n_q(t_\tau),
\end{equation}

where $n_q(t)$ is the number of tweets on topic $q$ in time interval $t$. We then calculate the ratios
$C_q(t_i, t_j) = N_q(t_i) / N_q(t_j)$ for topic $q$ for time frames $t_i$ and
$t_j$.  

\begin{figure} [ht]
\centering
\includegraphics[width=80mm, height=40mm] {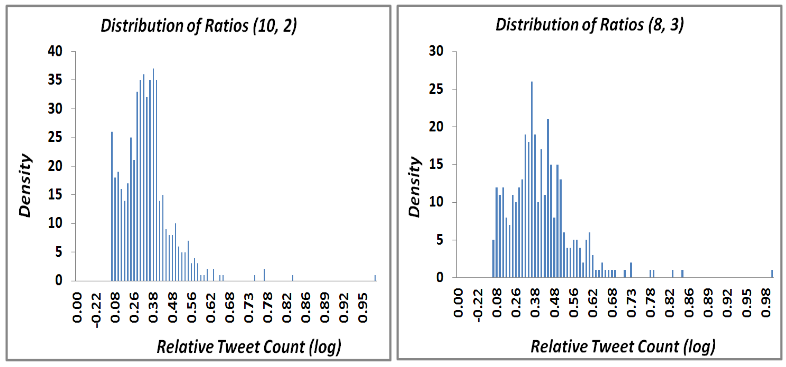}
\caption{The distribution of $C_q(t_i, t_j)$'s over all topics for two arbitrarily chosen pairs of time frames: (10, 2) and (8, 3)} \label{ratio_general}
\end{figure}

Figure \ref{ratio_general} shows the distribution of $C_q(t_i, t_j)$'s over all topics for two arbitrarily chosen pairs of time frames: (10, 2) and (8, 3) (nevertheless such that $t_i > t_j$, and $t_i$ is relatively large, and $t_j$ is small).

These figures suggest that the ratios $C_q(t_i, t_j)$ are distributed according to the log-normal distributions. We tested and confirmed that the distributions indeed follows the log-normal distributions.  

This finding agrees with the result from a similar experiment on Twitter trends. Asur et al \cite{Asur2011} argued that the log-normal distribution occurs due to the multiplicative process involved in the growth of trends which incorporates the decay of novelty as well as the rate of propagation. The intuitive explanation is that at each time step the number of new tweets (original tweets or retweets) on a topic is multiplied over the tweets that we already have. The number of past tweets, in turn, is a proxy for the number of users that are aware of the topic up to that point. These  users  discuss the topic on different forums, including Twitter, essentially creating an effective network through which the topic spreads. As more users talk about a particular topic, many others are likely to learn about it, thus giving the multiplicative nature of the spreading.  On the other hand, the monotically decreasing decaying process characterizes the decay in timeliness and novelty of the topic as it slowly becomes obsolete.

However, while only  35\% of the tweets in Twitter are retweets \cite{Asur2011}, there is a much larger percentage of tweets that are retweets in Sina Weibo. From our sample we observed that a high 65\% of the tweets are retweets. Thus, we can hypothesize that the effect of retweets is much larger in Sina Weibo. Sina Weibo users are more likely to learn about a particular topic through retweets. 

\subsection{The Evolution of Retweets and Original Tweets}

\begin{figure} [ht]
\centering
\includegraphics[width=85mm, height=40mm] {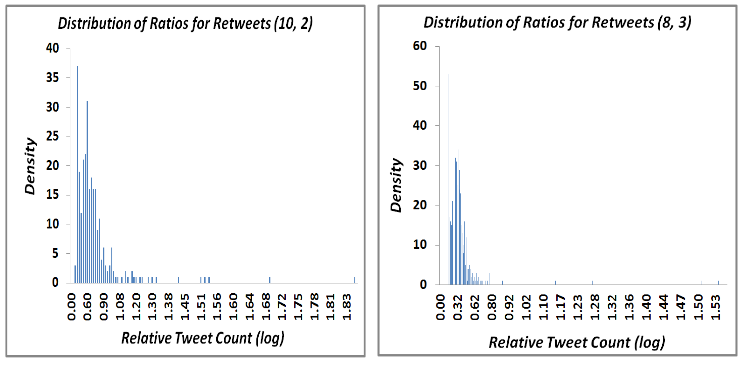}
\caption{The densities of ratios between cumulative retweets counts measured in two respective time frames} \label{ratio_retweet}
\end{figure}

\begin{figure} [ht]
\centering
\includegraphics[width=85mm, height=40mm] {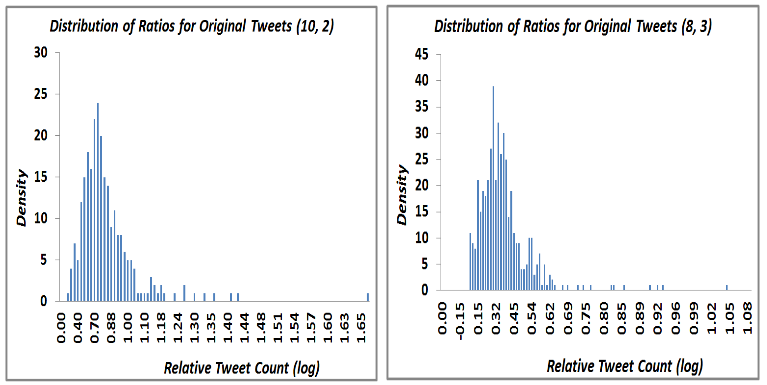}
\caption{The densities of ratios between cumulative original tweets counts measured in two respective time frames} \label{ratio_single}
\end{figure}

We separate the tweets in Sina Weibo into original tweets and retweets and calculated the densities of ratios between cumulative retweets/original tweets counts measured in two respective time frames. Figure \ref{ratio_retweet}  and Figure \ref{ratio_single} show the distribution of original tweets/retweets ratio over all topics for the same two pairs of time frames: (10, 2) and (8, 3)

We see from Figure \ref{ratio_single} that the distribution of original tweets ratios follows the log-normal distribution. However, from Figure \ref{ratio_retweet} we observe that the distribution of retweets ratios seems to leaning more towards the left. That is, there seems to be a large amount of of low retweet ratios in the distribution.  What's more, there are some extreme spikes in the lower ratios area of the distribution. 

We tested and confirmed that the distributions in Figure \ref{ratio_single} indeed correspond to log-normal distribution. However, we observe that the distribution of retweet ratios (Figure \ref{ratio_retweet}) does not satisfy all the properties of the log-normal distribution.

\subsection{ Identifying Top Retweeters in Sina Weibo}

\begin{figure} [ht]
\centering
\includegraphics[width=85mm, height=90mm] {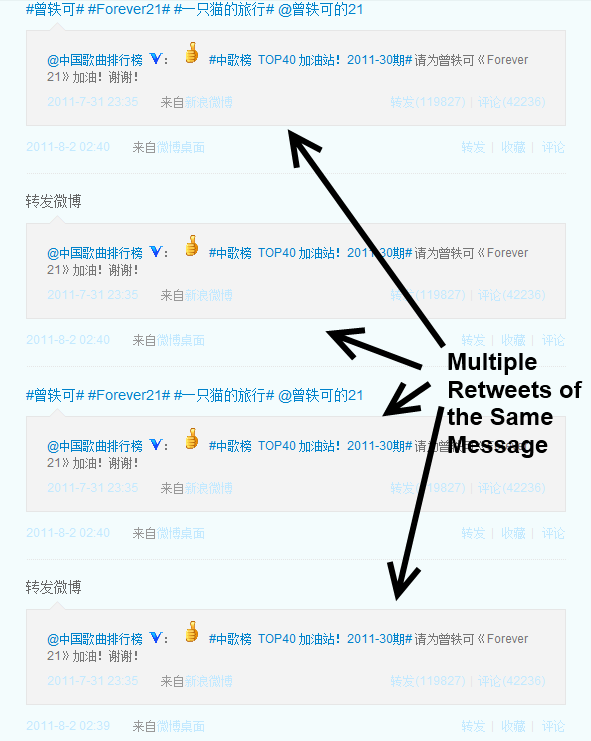}
\caption{An example of the activities of a spam account on Sina Weibo } \label{example_spam}
\end{figure}

From Figure \ref{ratio_retweet} in the previous Section we observed that there is a high percentage of low ratios in the distribution of retweet ratios. This means that for a lot of the topics, during the observed time duration the growth of retweets seems to be slower than usual. We hypothesize that this is due to the activities of certain users in Sina Weibo: as these accounts post a tweet, they tend to set up many other fake accounts to continuously retweet this tweet, expecting that the high retweet numbers would propel the tweet to place in the Sina Weibo hourly trending list. This would then cause other users to notice the tweet more after it has emerged as the top hourly, daily, or weekly trend setter.  Figure \ref{example_spam} illustrates the activity of a suspected spam account. We observe that it tend to continuously retweets the same posts from the same users. In fact, this particular account only retweets posts from one user. And, every post from that user gets retweeted multiple times. 

\begin{figure} [ht]
\centering
\includegraphics[width=85mm, height=40mm] {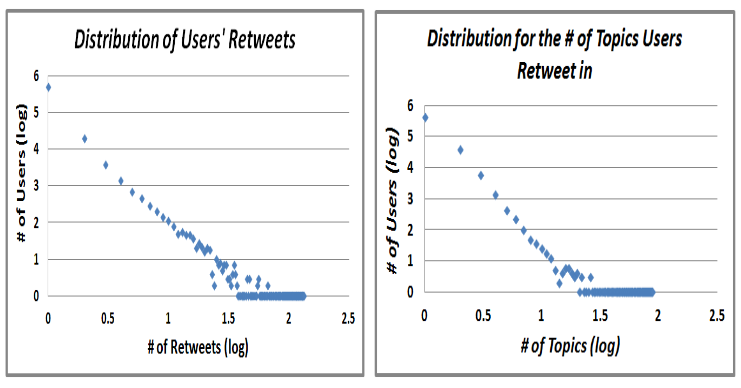}
\caption{The distribution for the number of users' retweets and the number of topics users' retweets trend in } \label{distribution_r_r}
\end{figure}

We attempt to verify the above hypothesis empirically. 
If the above hypothesis is true, we would expect that the distribution for the number of users' retweets to be scale free, since an abnormally large number of retweets would be by the supposed spammers. Figure \ref{distribution_r_r} a) illustrates the distribution for the number of users and their corresponding number of retweets (over all topics). Figure \ref{distribution_r_r} b) illustrates the distribution for the number of users and the numbers of topics that they caused to trend by their retweets. We observe that both distributions in Figure \ref{distribution_r_r}  follows the power law. 

\subsection{Identifying Spammers in Sina Weibo}

We define a spamming account as ones that are  set up for the purpose of repeatedly retweeting certain messages, thus giving these messages artificially inflated popularity. According to our hypothesis in the previous section,  the users who retweet abnormally high amounts are more likely to be spam accounts. We test this hypothesis by manually checking the top 40 accounts who retweeted the most. To our surprise, after one month, 37 of these 40 accounts can no longer be accessed. That is, after we enter the accounts' IDs, we retrieved a message from Sina Weibo stating that the account has been removed and can no longer be accessed. 



According to the Sina Weibo frequently asked questions page, such error pages are displayed after Sina Weibo administrators delete a questionable account. Sina Weibo users themselves can not delete or close down their own account. In order to do so, they have to contact the administrators at Sina Weibo for help. We assume that the only reason we can no longer retrieve an active account after one month is that this account was deleted by the administrator for performing malicious activities such as spamming, or spreading illegal information.  What's more, according to Sina Weibo's frequently asked question page, if a user sends a tweet containing illegal or sensitive information, such tweet will be immediately deleted by Sina Weibo's administrators, however, the users' accounts will still be active. For the above reason we assume that if an account was active one month ago and can no longer be reached, this account has very likely performed malicious activities such as spamming.

We inspect the user accounts with the most retweets in our sample and the number of accounts they retweeted.  We see that although the number of times these accounts retweeted was very high, they mostly only retweet messages from a few users. We re-organize the users who retweeted by the ratio between the number of times he/she retweeted and the number of users he/she retweeted. We refer to this as the user-retweet ratio. Table \ref{verified} illustrates the top 10 users with the highest user-retweet ratios.  We note that for all these users, they  each retweet posts from only one account. We observe that this is true for the top 30 accounts with the highest user-retweet ratios. 

\begin{table}[ht]
\centering
\begin{tabular}{|c|c|c|c|c|}
\hline
\small{User ID}& \small{\# Retweets} & \small{\#  Retweeted}&\small{U-R Ratio} \\
\hline
\small{1840241580}	&134		&1&134\\
\small{2241506824}	&125	&1&125\\		
\small{1840263604}	&68		&1&68	\\		
\small{1840237192}	&64	&1&64		\\		
\small{1840251632}	&64		&1&64	\\		
\small{2208320854}	&55		&1& 55	\\		
\small{2208320990}	&51	&1&51		\\		
\small{2208329370}	&48		&1&48	\\		
\small{2218142513}	&47		&1&47	\\		
\small{1843422117}	&44		&1&44	\\			
\hline
\end{tabular}
\caption{The top 10 accounts with the highest user-retweet ratios (u-r ratio)}
\label{verified}
\end{table}

We conduct the following experiment: starting from the users with the highest user-retweet ratios, we used a crawler to automatically visit and retrieve each user's Sina Weibo account. Thus we measured the percentage of user accounts that can still be accessed (as opposed to be directed to the error page) organized by user-retweet ratios  (Table \ref{still}).  We observe that only 12\% of the accounts with user-retweet ratios of above 30 are active. And, 
as user-retweet ratios decrease, the percentages of active accounts slowly increase. We consider this to be strong evidence for the hypothesis that user accounts with high user-retweet ratios are likely to be spam accounts.  

\begin{table}[ht]
\centering
\begin{tabular}{|c|c|c|}

\hline
Ratio&	\% Active Accounts&		\% Inactive Accounts\\	
\hline		
$\geq$30	&12\%&	88\%\\			
20 -- 29	&38\%	&63\%\\			
11 -- 19&	16\%&	84\%\\			
10&	22\%&		78\%\\			
9&	12\%&	88\%\\			
8&	16\%&	84\%\\			
7&	15\%&	85\%\\			
6&	21\%&	79\%\\			
5&	30\%&	70\%\\			
4&	58\%&		42\%\\			
3&	80\%&	20\%\\			
2&	96\%&	4\%\\
1& 92\% & 8\%\\		
\hline
\end{tabular}
\caption{The percentage of accounts whose profiles can still be accessed, organized by user-retweet ratio}
\label{still}
\end{table}

We observe that in some cases, accounts with lower user-retweet ratios can still be a spam account. For example, an account could retweet a number of posts from other spam accounts, thus minimizing the suspicion of being detected as a spam account itself.  

\subsection{Removing Spammers in Sina Weibo}

We hypothesize that the accounts deleted by the Sina Weibo administrator are mostly spam accounts. We identified these accounts in the previous section.

From our sample, after automatically checking each account,  we identified 4985 accounts that were deleted by the Sina Weibo administrator.  We called these 4985 accounts ``suspected spam accounts''. The total number of users that published tweets in our sample is 463,231, and the number of users who retweeted at least once in our sample is 187,130. Thus we identified 1.08\% of the total users (2.66\% of users that retweeted) as suspected spam accounts.

Next, in order to measure the effect of spam, we removed all retweets from our sample disseminated by suspected spam accounts as well as posts published by them (and then later retweeted by others).  We hypothesize that by removing these retweets, we can eliminate the influences caused by the suspected spam accounts. 

We observed that after these posts were removed, we were left with only 189,686 retweets in our sample (51\% of the original total retweets). In other words, by removing retweets associated wth suspected spam accounts, we successfully removed 182,429 retweets, which is 49\% of the total retweets and 32\% of total tweets (both retweets and original tweets) from our sample.  This result is very interesting because it shows that a large amount of retweets in our sample are associated with suspected spam accounts. The spam accounts are therefore artificially inflating the popularity of topics, causing them to trend. 

\begin{figure} [ht]
\centering
\includegraphics[width=85mm, height=50mm] {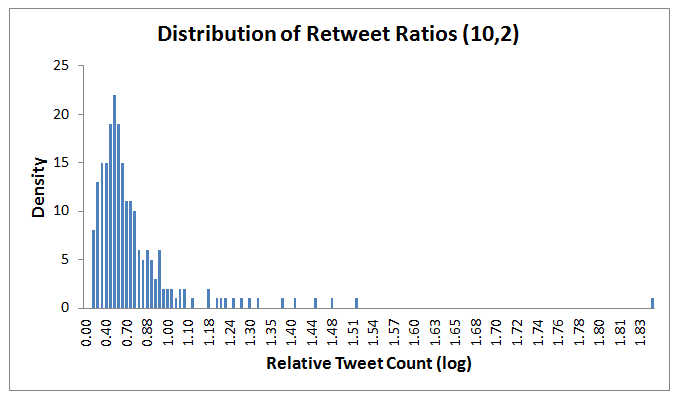}
\caption{The distribution of retweet ratios for time frame (10, 2) after the removal of tweets associated with suspected spam accounts } \label{correct}
\end{figure}

To see the difference after the posts associated with suspected spam accounts were removed, we re-calculated the distribution of user-retweet ratios again for arbitrarily chosen pairs of time frames. Figure \ref{correct} illustrates the distribution for time frames (10, 2). We observed that the distribution is now much smoother and seem to follow the log-normal distribution. We performed the log-normal test and verified that this is indeed the case.

\subsection{Spammers and Trend-setters}

We found 6824 users in our sample whose tweets were retweeted. We note that the total number of users who retweeted at least one person's tweet was 187130, however, these users were retweeting posts from only 6824 users. 

\begin{figure} [ht]
\centering
\includegraphics[width=85mm, height=50mm] {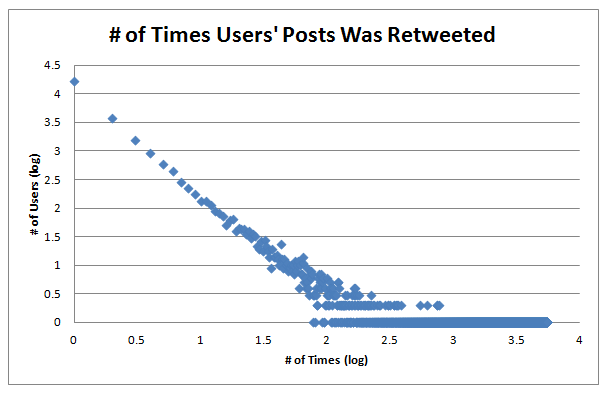}
\caption{The distribution for the \# of Times Users' Posts Were Retweeted } \label{distribution_r}
\end{figure}

Figure \ref{distribution_r} illustrates the distribution for the number of times users' posts were retweeted. We found that the distribution follows the power law.

Next, we further inspected the retweets in our sample associated with suspected spam accounts. We discovered that the number of users whose tweets were retweeted by the suspected spam accounts was 4665, which is a surprising {\bf 68\%} of the users whose posts were retweeted in our sample. This shows that the suspected spam accounts affect a majority of the trend-setters in our sample, helping them raise the retweet number of their posts and thereby making their posts appear on the trending list. The overall effect of the spammers is very significant. We also observed that a high 98\% of the total trending keywords can be found in posts retweeted by suspected spam accounts. Thus it can also be argued that many of the trends themselves are artificially generated, which is a very important result.

\begin{table*}[ht]
\centering
\begin{tabular}{|c|c|c|c|}

\hline
ID&	\# Times Posts Retweeted& Description &Verified\\
\hline		
1713926427&	5490&Silly Jokes & No\\
1843443790&	5466&Good Movies & No\\
1760945071&	3852& Chinese Groupon (tuan88.com) & Yes\\
1660209951&	3606& Global Fashion & No\\
1397618027&	3008& Comics & No\\
1879349260&	2954& Silly Stories & No\\
1854743504&	2789 & Hot Movies & No\\
1760717745&	2541&Weibo Horoscope &No\\
1642088277&	2087&Financial Magazine (caijing.com.cn)&Yes\\
2019719255&	2042	&Japanese Soap Operas & No\\	
\hline
\end{tabular}
\caption{Top 10 Users Whose Posts Are Most Retweeted}
\label{last_table}
\end{table*}

For the 4665 accounts whose tweets were retweeted by suspected spam accounts, we used a crawler to automatically inspect the types of accounts they are. We found that a high 79\% of the accounts are verified accounts, 5\% are expert accounts (see Section 3.2.1 for definitions of verified accounts and expert accounts). We manually inspected 50 of the verified accounts and found that they are mostly accounts with commercial purposes (such as the Sina Weibo accounts for travel agencies, super markets, fashion brands or hospitals).

We re-organized the users whose posts were retweeted by the number of times it was retweeted. Table \ref{last_table} illustrates the top 10 such users and the descriptions of their accounts. We observe that  only 2 out of the top 10 accounts were verified accounts.  9 out of 10 accounts seem to have a strong focus on collecting user-contributed jokes, movie trivia, quizzes, stories and so on. These accounts seem to operate as discussion and sharing platforms. The users who follow these accounts tend to contribute jokes or stories. Once they are posted, other followers tend to retweet them frequently. This result verifies a similar finding from Yu et al. \cite{Weibo}

\section{Conclusion and Future Work}

We have examined the tweets relating to the trending topics in Sina Weibo. First we analyze the growth and persistance of trends. 
When we looked at the distribution of tweets over time, we observed that there was a difference when contrasted with Twitter. 
The main reason for the difference was that the effect of retweets on Sina Weibo was significantly higher than on Twitter. 
We also found  (as our previous work~\cite{Weibo} suggests) that many of the accounts that contribute to trends tend to operate as user contributed online magazines, sharing amusing pictures, stories and antidotes. Such posts tend to recieve a large amount of responses from users and thus retweets.  

When we examined the retweets in more detail, we made an important discovery. We found that 49\% of the retweets in Sina Weibo containing trending keywords  were actually associated with fraudulent accounts. We observed that these accounts comprised of a small amount (1.08\% of the total users) of users but were responsible for a large percentage of the total retweets for the trending keywords. These fake accounts are responsible for artificially inflating certain posts, thus creating fake trends in Sina Weibo. 

We relate our finding to the questions we raised in the introduction. There is a strong competition among content in online social media to become popular and trend and this gives motivation to users to artificially inflate topics to gain a competitive edge. We hypothesize that certain accounts in Sina Weibo employ fake accounts to repeatedly repeat their tweets in order to propel them to the top trending list, thus gaining prominence as top trend setters (and more visible to other users). We found evidence suggesting that the accounts that do so tend to be verified accounts with commercial purposes. 

In the future, we would like to examine the behavior of these fake accounts that contribute to artificial inflation in Sina Weibo to learn how successful they are in influencing trends. 
\bibliographystyle{siam}

\end{document}